\input harvmac.tex
\input epsf.tex
\parindent=0pt
\parskip=5pt

\hyphenation{satisfying}

\def\IR{{\hbox{{\rm I}\kern-.2em\hbox{\rm R}}}}
\def\IB{{\hbox{{\rm I}\kern-.2em\hbox{\rm B}}}}
\def\IN{{\hbox{{\rm I}\kern-.2em\hbox{\rm N}}}}
\def\IC{\,\,{\hbox{{\rm I}\kern-.59em\hbox{\bf C}}}}
\def\IZ{{\hbox{{\rm Z}\kern-.4em\hbox{\rm Z}}}}
\def\IP{{\hbox{{\rm I}\kern-.2em\hbox{\rm P}}}}
\def\IH{{\hbox{{\rm I}\kern-.4em\hbox{\rm H}}}}
\def\ID{{\hbox{{\rm I}\kern-.2em\hbox{\rm D}}}}
\def\II{{\hbox{\rm I}\kern-.2em\hbox{\rm I}}}

\noblackbox

\leftline{\epsfxsize1.0in\epsfbox{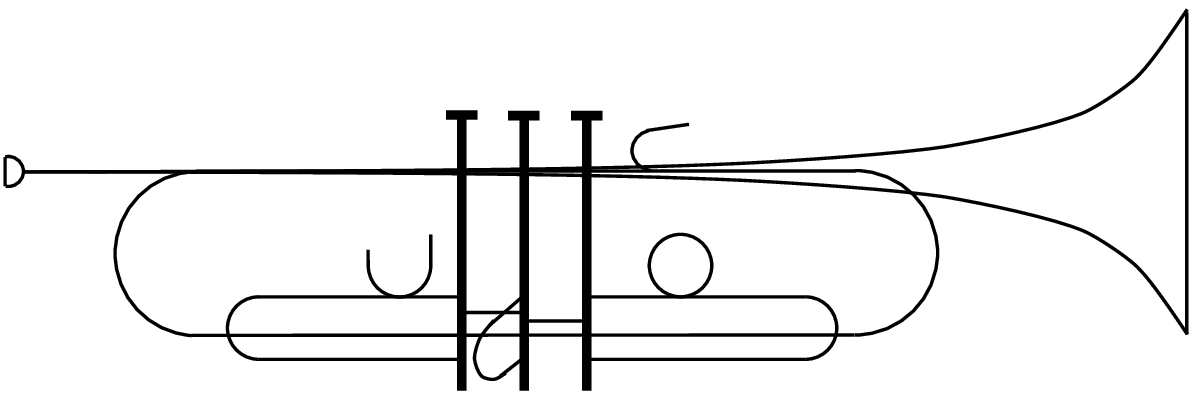}}
\vskip-1.0cm
\Title{\hbox{hep-th/9805047}}
{More Superstrings from Supergravity}

\centerline{\bf Clifford V. Johnson$^\flat$}

\bigskip
\bigskip

\vbox{\baselineskip12pt\centerline{\hbox{\it 
Department of Physics}}
\centerline{\hbox{\it University of California}}
\centerline{\hbox{\it Santa Barbara CA 93106, U.S.A.}}}

\footnote{}{\sl email: $^\flat${\tt cvj@itp.ucsb.edu}$\,\,$
On leave from the Dept. of
Physics and Astronomy, University of Kentucky, Lexington KY~40502,
U.S.A.}

\vskip1.0cm
\centerline{\bf Abstract}
\vskip0.7cm
\vbox{\narrower\baselineskip=12pt\noindent
The four six--dimensional ``little string'' theories are all described
in the infinite momentum frame (IMF) as matrix theories by
non--trivial 1+1 dimensional infra--red fixed points.  We characterize
these fixed points using supergravity. Starting from the matrix theory
definition of M5--branes, we derive an associated dual supergravity
description of the fixed point theories, arising as the near horizon
geometry of certain brane configurations. These supergravity solutions
are all smooth, and involve three dimensional Anti--de Sitter space
$AdS_3$. They therefore provide a complete description of the fixed
point theories, and hence the IMF little string theories, if the
AdS/CFT correspondence holds.}

\Date{9th May 1998}
\baselineskip13pt
\lref\dbranes{J.~Dai, R.~G.~Leigh and J.~Polchinski, 
{\sl `New Connections Between String Theories'}, Mod.~Phys.~Lett.
{\bf A4} (1989) 2073\semi P.~Ho\u{r}ava, {\sl `Background Duality of
Open String Models'}, Phys. Lett. {\bf B231} (1989) 251\semi
R.~G.~Leigh, {\sl `Dirac--Born--Infeld Action from Dirichlet Sigma
Model'}, Mod.~Phys.~Lett. {\bf A4} (1989) 2767\semi J.~Polchinski,
{\sl `Combinatorics Of Boundaries in String Theory'}, Phys.~Rev.~D50
(1994) 6041, hep-th/9407031.}

\lref\orientifolds{A. Sagnotti, in {\sl `Non--Perturbative Quantum
 Field Theory'}, Eds. G. Mack {\it et. al.} (Pergammon Press, 1988),
p521\semi V. Periwal, unpublished\semi J. Govaerts, Phys. Lett. {\bf
B220} (1989) 77\semi P. Hor\u{a}va, {\sl `Strings on World Sheet
Orbifolds'}, Nucl. Phys. {\bf B327} (1989) 461.}
\lref\nsfivebrane{A. Strominger,  {\sl `Heterotic Solitons'} Nucl. Phys. 
{\bf B343}, (1990) 167; 
{\it Erratum: ibid.}, {\bf 353} (1991) 565.}
\lref\robin{R. W. Allen, I. Jack and D. R. T. Jones, 
{\sl `Chiral Sigma Models and the Dilaton Beta Function'}, 
Z. Phys. C {\bf 41} 
(1988) 323.}

\lref\sethi{S. Sethi, {\sl `The Matrix Formulation of Type IIB Five--Branes'},
 hep-th/9710005\semi A. Kapustin and S. Sethi, {\sl`The Higgs Branch
of Impurity Theories'}, hep-th/9804027.}

\lref\banksreview{For a review, with references,
see {\sl `Matrix Theory'}, T. Banks, hep-th/9710231.}
\lref\rutgers{D-E. Diaconescu, J. Gomis, 
{\sl `Matrix Description of Heterotic Theory on K3'}, hep-th/9711105.}
\lref\sen{A. Sen,  {\sl `D0--Branes on $T^n$ and Matrix Theory'}, 
hep-th/9709220.}
\lref\seibergtwo{N. Seiberg, 
{\sl `Why is the Matrix Model Correct?'}, Phys. Rev. Lett. {\bf 79}
 (1997) 3577, hep-th/9710009.}
\lref\seibergthree{N. Seiberg, {\sl `Notes on Theories with 16 
Supercharges'},  hep-th/9705117.}
\lref\paul{ P. S. Aspinwall, {\sl `Some Relationships Between Dualities in 
String Theory'}, Nucl. Phys. Proc. Suppl. {\bf 46} (1996) 30,
hep-th/9508154.}
\lref\john{J. H. Schwarz, 
{\sl `The Power of M Theory'}, Phys. Lett. {\bf B367} (1996) 97,
 hep-th/9510086.}
\lref\juan{J. Maldacena, {\sl `The Large $N$ Limit of Superconformal Field
 Theories and Supergravity'}, hep-th/9711200.}
\lref\juantwo{N. Itzhaki, J. M. Maldacena, J. Sonnenschein 
and S. Yankielowicz, {\sl `Supergravity and The Large N Limit of Theories
With Sixteen Supercharges'}, hep-th/9802042.}
\lref\correspond{S. S. Gubser, I. R. Klebanov, A. M. Polyakov,
 {\sl `Gauge Theory Correlators from Non-Critical String Theory'},
 hep-th/9802109\semi E. Witten, {\sl `Anti De Sitter Space And
 Holography'}, hep-th/9802150.}
\lref\ofer{O. Aharony, Y. Oz and  Z. Yin,
 {\sl `M--Theory on $AdS_p{\times}S^{11-p}$ and Superconformal Field
 Theories'}, hep-th/9803051.}
\lref\vafa{C. Vafa, {\sl `Evidence  for F--Theory'}, Nucl. Phys.
 {\bf B469} 403 (1996),  hep-th/9602022.}
\lref\amied{A. Hanany and E. Witten, {\sl `Type IIB Superstrings, 
BPS Monopoles, And Three-Dimensional Gauge Dynamics'}, Nucl. Phys. B492
(1997) 152-190, hep-th/9611230.}
\lref\ofereva{ O. Aharony, M. Berkooz, S. Kachru, N. Seiberg 
and E. Silverstein, {\sl `Matrix Description of Interacting Theories
in Six Dimensions'}, Adv. Theor. Math. Phys. {\bf 1} (1998) 148-157,
hep-th/9707079.}
\lref\oferseiberg{O. Aharony, M. Berkooz and N. Seiberg, 
{\sl `Light--Cone Description of (2,0) Superconformal Theories in Six
Dimensions'}, hep-th/9712117.}
\lref\edhiggs{E. Witten, {\sl `On The Conformal Field Theory of
 The Higgs Branch'}, JHEP 07 (1997) 003,  hep-th/9707093.}
\lref\stromingervafa{A. Strominger and C. Vafa, {\sl `Microscopic 
Origin of the Bekenstein--Hawking Entropy'}, Phys. Lett. {\bf B379}
(1996) 99, hep-th/9601029.}
\lref\douglasetal{M. R. Douglas, J. Polchinski and A Strominger, 
{\sl `Probing Five-Dimensional Black Holes with D-Branes'}, JHEP
12 (1997) 003, hep-th/9703031.}
\lref\gregsamson{A. Losev, G. Moore and  S. L. Shatashvili,
 {\sl `M\&m's'}, hep-th/9707250.}
\lref\dvviv{R. Dijkgraaf, E. Verlinde and H. Verlinde, 
{\sl `Notes on Matrix and Micro Strings'}, hep-th/9709107.}

\lref\berkooz{M. Berkooz and M. R. Douglas, 
{\sl `Five-branes in M(atrix) Theory'}, Phys. Lett. {\bf B395} (1997)
196, hep-th/9610236.}
\lref\dvviii{R. Dijkgraaf, E. Verlinde and H. Verlinde, 
{\sl `BPS Quantization of the Five-Brane'}, Nucl. Phys. {\bf B486}
(1997) 89, hep-th/9604055; {\it ibid.}, {\sl `BPS Spectrum of the
Five-Brane and Black Hole Entropy'}, Nucl. Phys. {\bf B486} (1997) 77,
hep-th/9603126; {\it ibid.}, {\sl
`Counting Dyons in $N{=}4$ String Theory'}, Nucl. Phys. {\bf B484} (1997)
543, hep-th/9607026\semi R. Dijkgraaf, G. Moore, E. Verlinde and
H. Verlinde, {\sl `Elliptic Genera of Symmetric Products and Second
Quantized Strings'}, Commun. Math. Phys. {\bf 185} (1997) 197,
hep-th/9608096.}
\lref\juanandy{J. Maldacena and A. Strominger, {\sl`$AdS_3$ 
Black Holes and a Stringy Exclusion Principle'}, hep-th/9804085.}
\lref\martinec{E. Martinec, {\sl`Matrix Models of AdS Gravity'}, 
hep-th/9804167.}
\lref\boon{H. J. Boonstra, B. Peeters and  K. Skenderis, 
{\sl`Brane intersections, anti-de Sitter spacetimes and dual
superconformal theories'}, hep-th/9803231.}
\lref\deger{S. Deger, A. Kaya, E. Sezgin, P. Sundell,
 {\sl `Spectrum of $D{=}6$, $N{=}4b$ Supergravity on
 $AdS_3{\times}S^3$'}, hep-th/9804166.}
\lref\jan{J. de Boer, unpublished.}
\lref\vafaii{C. Vafa, {\sl `Puzzles at Large $N$'}, hep-th/9804172.}
\lref\mads{H. J. Boonstra, B. Peeters and  K. Skenderis, 
{\sl `Duality and asymptotic geometries'}, Phys. Lett. {\bf B411}
(1997) 59, hep-th/9706192.}
\lref\paulii{P. S. Aspinwall, {\sl `M--Theory Versus 
F--Theory Pictures of the Heterotic String'},
Adv. Theor. Math. Phys. {\bf 1} (1998) 127, hep-th/9707014.}
\lref\harveystrominger{J. A. Harvey and A. Strominger,
 {\sl `The Heterotic String is a Soliton'}, Nucl. Phys. {\bf B449}
 (1995) 535, hep-th/9504047.}
\lref\sendual{A. Sen, {\sl `String String Duality Conjecture in 
Six Dimensions and Charged Solitonic Strings'}, Nucl. Phys. B450
(1995) 103, hep-th/9504027.}
\lref\andygaryjuan{G. Horowitz, J. M Maldacena and A. Strominger, 
{\sl `Nonextremal Black Hole Microstates and U--duality'},
Phys.Lett. B383 (1996) 151, hep-th/9603109.}
\lref\tseytlin{A. Tseytlin, {\sl `Harmonic superpositions of M--branes'}, 
Nucl. Phys. {\bf B475} (1996) 149, hep-th/9604035.}
\lref\gauntlett{J. P. Gauntlett, 
D. A. Kastor and J. Traschen, {\sl `Overlapping Branes in M--theory'},
Nucl. Phys. {\bf B478} (1996) 544, hep-th/9604179.}
\lref\martinecii{M. Li, and E. Martinec, 
{\sl `On the Entropy of Matrix Black Holes'}, Class. Quant. Grav. {\bf
 14} (1997) 3205, hep-th/9704134.}
\lref\juaniv{J. M. Maldacena, {\sl `Statistical Entropy of Near Extremal 
Five--branes'}, Nucl. Phys. B477 (1996) 168, hep-th/9605016.}

\lref\sethisuss{S. Sethi and L. Susskind, {\sl `Rotational 
Invariance in the M(atrix) Formulation of Type IIB Theory'},
 Phys. Lett. {\bf B400} (1997) 265, hep-th/9702101.}

\lref\heteroticcosets{C. V. Johnson, {\sl `Heterotic Coset Models'},
 Mod. Phys. Lett. {\bf A10} (1995) 549, hep-th/9409062\semi {\it
ibid.,} {\sl `Exact Models of Extremal Dyonic 4D Black Hole Solutions
of Heterotic String Theory'}, Phys. Rev.  {\bf D50} (1994) 4032,
hep-th/9403192.}

\lref\diacon{D.--E. Diaconescu and N. Seiberg, 
{\sl `The Coulomb Branch of $(4,4)$ Supersymmetric Field Theories in 
Two Dimensions'}, JHEP, {\bf 07}
 (1997) 001, hep-th/9707158.}
\lref\mikejoeandy{M. R. Douglas, J. Polchinski, and A. Strominger, 
{\sl`Probing
 Five Dimensional Black Holes with D--branes'}, hep-th/9703031.}
\lref\callan{C. G. Callan, J.A. Harvey and A. Strominger, 
{\sl `Supersymmetric String Solitons'}, in
Trieste 1991, proceedings, ``String Theory and Quantum Gravity'',
hep-th/9112030.}
\lref\sjrey{S--J. Rey, in
 {\sl `Superstrings and Particle Theory: Proceedings'}, edited by
   L. Clavelli and B. Harms, (World Scientific, 1990) 291\semi
   S--J. Rey, {\sl `The Confining Phase of Superstrings and Axion Strings'},
 Phys. Rev. {\bf D43} (1991) 526\semi I. Antoniades,
   C. Bachas, J. Ellis and D. Nanopoulos, {\sl `Cosmological String
   Theories and Discrete Inflation'}, Phys. Lett.  {\bf B211} (1988)
   393\semi {\it ibid.,} {\sl `An Expanding Universe in String
   Theory'}, Nucl. Phys. {\bf 328} (1989) 117.}
\lref\robdilaton{R.~C.~Myers, {\sl `New Dimensions for Old Strings'}, 
Phys. Lett. {\bf B199} (1987) 371.}
\lref\ff{B. L. Feigin and D. B. Fuchs, Funct. Anal. Appl. {\bf 16} (1982)
 114, {\it ibid.}, {\bf 17} (1983) 241.}
\lref\gepner{D. Gepner, {\sl `Spacetime Supersymmetry in Compactified 
String Theory and Superconformal Models'}, Nucl. Phys. {\bf B296} (1988) 757.}

\lref\mtwo{E. Bergshoeff, E. Sezgin and P. K. Townsend,
 {\sl `Supermembranes and Eleven Dimensional Supergravity'}, Phys.
Lett. {\bf B189} (1987) 75\semi M. J. Duff and K. S. Stelle, {\sl
`Multimembrane solutions of D=11 Supergravity'}, Phys. Lett. {\bf
B253} (1991) 113.}
\lref\mfive{R. G\"uven, {\sl `Black $p$--Brane 
Solutions of D=11 Supergravity theory'}, Phys. Lett. {\bf B276} (1992)
49.}

\lref\town{P. Townsend, {\sl `The eleven-dimensional 
supermembrane revisited'},
  Phys. Lett. {\bf B350} (1995) 184, hep-th/9501068.}
\lref\goed{E. Witten, {\sl `String Theory Dynamics in Various Dimensions'}, 
Nucl. Phys. {\bf B443} (1995) hep-th/9503124.}
\lref\wznw{S. P. Novikov, Ups. Mat. Nauk. {\bf 37} (1982) 3\semi
E. Witten, {\sl `Non--Abelian Bosonization in Two Dimensions'},
Comm. Math. Phys. {\bf 92} (1984) 455.}
\lref\rohm{R. Rohm, {\sl `Anomalous Interactions for the Supersymmetric 
Non--linear Sigma Model in Two Dimensions'}, Phys. Rev. {\bf D32} (1984) 2849.}

\lref\romans{L. Romans, {\sl `Massive $N{=}2A$ 
Supergravity In Ten-Dimensions'}, Phys. Lett. {\bf B169} (1986) 374.}
\lref\others{E. Bergshoeff, M. de Roo, M. B. Green, G. Papadopoulos
 and P. K. Townsend, 
{\sl `Duality of Type II 7-branes and 8-branes'}, Nucl.Phys. {\bf
B470} (1996) 113, hep-th/9601150.}

\lref\duff{Duff, Minasian and Witten, {\sl `Evidence for Heterotic/Heterotic 
Duality'}, Nucl. Phys. {\bf B465} (1996) 413, hep-th/9601036.}
\lref\duffetal{M. J. Duff, {\sl 
`Strong/Weak Coupling Duality from the Dual String', Nucl. Phys. {\bf B442}
(1995) 47, hep-th/9501030.}\semi M. J. Duff, {\sl `Putting
string/string duality to the test', Nucl. Phys. {\bf B436} (1995) 507,
hep-th/9406198.}}
\lref\berkoozi{M. Berkooz, R. G. Leigh, J. Polchinski, J. Schwarz, N. Seiberg 
and E. Witten, {\sl `Anomalies, Dualities, and Topology of D=6 N=1 Superstring
Vacua'}, Nucl. Phys. {\bf B475} (1996) 115, hep-th/9605184.}
\lref\rozali{M. Rozali,
 {\sl `Matrix Theory and U--Duality in Seven Dimensions'}, 
Phys. Lett. {\bf B400} (1997) 260, hep-th/9702136.}
\lref\berkoozii{M. Berkooz, N. Seiberg and M. Rozali, {\sl 
`Matrix Description of 
M-theory on $T^4$ and $T^5$'}, Phys. Lett. {\bf B408}
 (1997) 105, hep-th/9704089.}
\lref\berkooziii{M. Berkooz, and M. Rozali, 
{\sl `String Dualities from Matrix Theory'}, hep-th/9705175.}

\lref\atish{A. Dahbolkar and J. Park, {\sl `Strings on Orientifolds'},
 Nucl. Phys.  {\bf B477} (1996) 701, hep-th/9604178.}
\lref\bfss{T. Banks, W. Fischler, S. Shenker and L. Susskind, {\sl 
 `M--Theory As A 
Matrix Model: A Conjecture'}, Phys. Rev. {\bf D55}
 (1997) 5112, hep-th/9610043.}
\lref\edjoe{J. Polchinksi and E. Witten,  {\sl 
`Evidence for Heterotic - Type I 
String Duality'}, Nucl. Phys. {\bf B460} (1996) 525, hep-th/9510169.}
\lref\seiberg{N. Seiberg, {\sl `Matrix Description of M-theory on $T^5$ and 
$T^5/Z_2$'}, Phys. Lett. {\bf B408} (1997) 98, hep-th/9705221.}
\lref\sagnotti{M. Bianchi and A. Sagnotti, {\sl `Twist
 Symmetry and Open String Wilson Lines'} Nucl. Phys. {\bf B361} (1991)
519.}
\lref\sagnottii{A. Sagnotti, {\sl 
`A Note on the Green - Schwarz Mechanism in Open - String Theories'},
Phys. Lett. {\bf B294} (1992) 196, hep-th/9210127.}
\lref\horavawitten{P. Horava and E. Witten, {\sl `Heterotic and Type I String 
Dynamics from Eleven Dimensions'}, Nucl. Phys. {\bf B460} (1996) 506,
hep-th/9510209.}

\lref\dvv{R. Dijkgraaf, E. Verlinde, H. Verlinde, 
{\sl `Matrix String Theory'}, Nucl. Phys. {\bf B500} (1997) 43,
hep-th/9703030.}
\lref\dvvii{R. Dijkgraaf, E. Verlinde, H. Verlinde, {\sl `5D Black Holes and 
Matrix Strings'}, Nucl. Phys. {\bf B506} 121 (1997),  hep-th/9704018.}
\lref\robme{C. V. Johnson and R. C. Myers, 
{\sl `Aspects of Type IIB Theory on Asymptotically Locally Euclidean
Spaces'}, Phys. Rev. {\bf D55} (1997) 6382, hep-th/9610140.}
\lref\douglasi{M. R. Douglas, {\sl `Branes within Branes'}, hep-th/9512077.}

\lref\mumford{D. Mumford and J. Fogarty, {\sl `Geometric Invariant Theory'},
Springer, 1982.}
\lref\edcomm{E. Witten, {\sl `Some Comments On String Dynamics'}, in the 
Proceedings of {\sl Strings 95}, USC, 1995, hep-th/9507121.}

\lref\joetensor{J. Polchinski, {\sl `Tensors From $K3$ Orientifolds'}, 
hep-th/9606165.}
\lref\kronheimer{P. B. Kronheimer, {\sl `The Construction of ALE Spaces as 
Hyper--K\"ahler Quotients'}, J.~Diff. Geom. {\bf 29} (1989) 665.}
\lref\hitchin{N. J.  Hitchin,  {\sl `Polygons and Gravitons'}, Math. Proc. 
Camb. Phil. Soc. {\bf 85} (1979) 465.}
\lref\douglasmoore{M. R. Douglas and G. Moore,
  {\sl `D--Branes, Quivers and ALE Instantons'}, hep-th/9603167.}

\lref\hitchinetal{N. J. Hitchin, A. Karlhede, U. Lindstr\"om and M. Ro\u cek, 
{\sl `Hyper--K\"ahler Metrics and Supersymmetry'}, Comm. Math. Phys. {\bf 108}
(1987) 535.}

\lref\klein{F. Klein, {\sl `Vorlesungen \"Uber das Ikosaeder und die 
Aufl\"osung der Gleichungen vom f\"unften Grade'}, Teubner, Leipzig 1884;
F. Klein, {\sl `Lectures on the Icosahedron and  the Solution of an Equation 
of Fifth Degree'}, Dover, New York, 1913.}

\lref\elliot{J. P. Elliot and P. G. Dawber, {\sl `Symmetry in Physics'}, 
McMillan, 1986.}

\lref\ericjoe{E. G. Gimon and J. Polchinski, {\sl `Consistency
 Conditions of Orientifolds and D--Manifolds'}, Phys. Rev. {\bf D54} (1996) 
1667, hep-th/9601038.}
\lref\ericmeI{E. G. Gimon and C. V. Johnson, {\sl `$K3$ Orientifolds'}, Nucl. 
Phys. {\bf B478} (1996), hep-th/9604129.}
\lref\ericmeII{E. G. Gimion and C. V. Johnson, {\sl `Multiple Realisations of 
${\cal N}{=}1$ Vacua in Six Dimensions'}, Nucl. Phys. {\bf B479} (1996), 285,
hep-th/9606176}
\lref\mackay{J. McKay, {\sl `Graphs, Singularties
 and Finite Groups'}, Proc. Symp. Pure. Math. {\bf 37} (1980) 183,
Providence, RI; Amer. Math. Soc.}

\lref\orbifold{L. Dixon, J. Harvey, C. Vafa and E. Witten, {\sl `Strings on 
Orbifolds'}, Nucl. Phys. {\bf B261} (1985) 678;
{\it ibid}, Nucl. Phys. {\bf B274} (1986) 285.}
\lref\algebra{P. Slodowy, {\sl `Simple Singularities and Simple Algebraic 
Groups'}, Lecture Notes in Math., Vol.  {\bf 815}, Springer, Berlin, 1980.}
\lref\gibhawk{G. W. Gibbons and S. W. Hawking, {\sl `Gravitational
Multi--Instantons'}, Phys. Lett. {\bf B78} (1978) 430.}
\lref\eguchihanson{T. Eguchi and A. J. Hanson, {\sl `Asymptotically Flat 
Self--Dual Solutions to Euclidean Gravity'}, Phys. Lett. {\bf B74} (1978) 249.}

\lref\wittenadhm{E. Witten, {\sl `Sigma Models and the ADHM Construction of 
Instantons'}, J.~Geom.  Phys. {\bf 15} (1995) 215, hep-th/9410052.}
\lref\edsmall{E. Witten, {\sl `Small Instantons in String Theory'},  Nucl. 
Phys. {\bf B460} (1996) 541, hep-th/9511030.}
\lref\douglasii{M. R.  Douglas, {\sl `Gauge Fields and D--Branes'},  
hep-th/9604198.}

\lref\phases{E. Witten, {\sl `Phases of $N{=}2$ Theories in Two Dimensions'}, 
Nucl. Phys. {\bf B403} (1993) 159,  hep-th/9301042.}
\lref\edbound{E. Witten, {\sl `Bound States of Strings and $p$--Branes'}, 
Nucl. Phys. {\bf B460} (1996) 335, hep-th/9510135.}

\lref\ADHM{M. F. Atiyah, V. Drinfeld, N. J. Hitchin and Y. I. Manin, {\sl
`Construction of Instantons'} Phys. Lett. {\bf A65} (1978) 185.}

\lref\kronheimernakajima{P. B. Kronheimer and H. Nakajima, {\sl `Yang--Mills 
Instantons on ALE Gravitational Instantons'}, Math. Ann. {\bf 288} (1990) 263.}
\lref\italiansi{M. Bianchi, F. Fucito, G. Rossi, and M. Martinelli, 
{\sl `Explicit Construction of Yang--Mills Instantons on ALE Spaces'}, 
Nucl. Phys. {\bf B473} (1996) 367, hep-th/9601162.}

\lref\italiansii{D. Anselmi, M. Bill\'o, P. Fr\'e, L. Giraradello and A. 
Zaffaroni, {\sl `ALE Manifolds and Conformal Field Theories'},  Int. J. 
Mod. Phys. {\bf A9} (1994) 3007,  hep-th/9304135.}

\lref\nonrenorm{L. Alvarez--Gaume and D. Z. Freedman, {\sl `Geometrical 
structure and Ultraviolet Finiteness in the Supersymmetric Sigma Model'}, 
Comm. Math. Phys. {\bf 80} (1981) 443.} 
\lref\myoldpaper{C. V. Johnson, {\sl `Exact Models of Extremal Dyonic 4D 
Black Hole Solutions of Heterotic String Theory'}, Phys. Rev. {\bf D50} (1994)
4032, hep-th/9403192.}

\lref\gojoe{J. Polchinski, {\sl `Dirichlet Branes and Ramond--Ramond Charges
 in String Theory'}, Phys. Rev. Lett. {\bf 75} (1995) hep-th/9510017.}
\lref\dnotes{J. Polchinski, S. Chaudhuri and C. V. Johnson, {\sl `Notes on 
D--Branes'}, hep-th/9602052.}

\lref\joetasi{J. Polchinski, `TASI Lectures on D-Branes', hep-th/9611050.}
\lref\hull{C. M.  Hull, {\sl `String--String Duality in Ten Dimensions'}, 
 Phys. Lett. {\bf B357} (1995) 545,  hep-th/9506194.}

\lref\dine{M. Dine, N. Seiberg and E. Witten, {\sl 
`Fayet--Iliopolos Terms in String Theory'}, Nucl. Phys. {\bf B289}
(1987) 589.}
\lref\taylor{W. Taylor,
{\sl `D--Brane field theory on compact spaces'}, Phys.Lett. B394
 (1997) 283, hep-th/9611042\semi O. J. Ganor, S. Ramgoolam, W. Taylor,
 {\sl `Branes, Fluxes and Duality in M(atrix)-Theory'},
 Nucl. Phys. {\bf B492} (1997) 191, hep-th/9611202.}

\lref\matrixeight{D. Lowe, {\sl `$E_8{\times}E_8$
 Instantons in Matrix Theory '}, hep-th/9709015\semi O. Aharony,
M. Berkooz, S. Kachru, and E. Silverstein, {\sl `Matrix Description of
$(1,0)$ Theories in Six Dimensions'}, Phys. Lett. {\bf B420} (1998)
55, hep-th/9709118.}
\lref\eva{S. Kachru and E. Silverstein, {\sl `On Gauge Bosons in
 the Matrix Model Approach to M~Theory'}, Phys. Lett. {\bf B396} (1997)
70, hep-th/9612162.}
\lref\rey{N. Kim  and S-J. Rey, {\sl `M(atrix) Theory on an Orbifold and 
Twisted Membrane'}, Nucl. Phys. {\bf B504} (1997) 189,
hep-th/9701139\semi S-J. Rey, {\sl `Heterotic M(atrix) Strings and
Their Interactions'}, Nucl. Phys. {\bf B502} 170,1997,
hep-th/9704158.}
\lref\banksmotl{T. Banks and L. Motl, 
{\sl `Heterotic Strings from Matrices'}, JHEP 12 (1997) 004, hep-th/9703218.}
\lref\matrixheterotic{L. Motl, {\sl `Quaternions and M(atrix) Theory
in Spaces with Boundaries'}, hep-th/9612198\semi T. Banks, N. Seiberg,
E. Silverstein, {\sl `Zero and One-dimensional Probes with N=8
Supersymmetry'}, Phys. Lett. {\bf B401} (1997) 30, hep-th/9703052\semi
D. Lowe, {\sl `Bound States of Type I' D-particles
and Enhanced Gauge Symmetry'}, Nucl. Phys. {\bf B501} (1997) 134,
hep-th/9702006\semi D. Lowe, {\sl `Heterotic Matrix String Theory'},
Phys. Lett. {\bf B403} (1997) 243, hep-th/9704041.}
\lref\petr{P. Horava, {\sl `Matrix Theory and
Heterotic Strings on Tori'}, Nucl.  Phys. {\bf B505} 84 (1997),
hep-th/9705055.}
\lref\kabat{D. Kabat and
S-J. Rey, {\sl `Wilson Lines and T-Duality in Heterotic M(atrix)
Theory'}, Nucl. Phys. {\bf B508} 535, (1997) hep-th/9707099.}
\lref\matrixheteroticii{S. Govindarajan, {\sl `Heterotic
M(atrix) theory at generic points in Narain moduli space'},
hep-th/9707164.}
\lref\motl{L. Motl, {\sl `Proposals on Non--Perturbative Superstring
 Interactions'}, hep-th/9701025.}
\lref\banks{T. Banks and N. Seiberg, {\sl `Strings from Matrices'},
 Nucl. Phys. {\bf B497} 41 (1997), hep-th/9702187.}
\lref\danielsson{U. H. Danielsson, G. Ferretti, {\sl `The Heterotic
 Life of the D-particle'}, Int. J. Mod. Phys. {\bf A12} 
(1997) 4581, hep-th/9610082.}
\lref\douglasooguri{M. R. Douglas, H. Ooguri, S. H. Shenker, 
{\sl `Issues in M(atrix) Theory Compactification'}, Phys.Lett. {\bf
 B402} (1997) 36, hep-th/9702203\semi M. R. Douglas, H. Ooguri {\sl
 `Why Matrix Theory is Hard'}, hep-th/9710178.}
\lref\fischler{W. Fischler, A. Rajaraman, 
{\sl `M(atrix) String Theory on K3'},
 hep-th/9704123.}
\lref\edmtheory{E. Witten, {\sl `Five-branes And 
$M$-Theory On An Orbifold'}, Nucl. Phys. {\bf B463} (1996) 383,
hep-th/9512219.}
\lref\sigmamodels{S. J. Gates, C. M. Hull and M. Ro\u{c}ek, 
{\sl `Twisted Multiplets and New Supersymmetric Non--Linear Sigma Models'}, 
Nucl. Phys. 
{\bf B248} (1984) 157\semi M. Ro\u{c}ek, K. Schoutens and A. Sevrin, 
{\sl `Off--Shell WZW Models in Extended Superspace'},
Phys. Lett. {\bf B265} (1991) 303\semi M. Ro\u{c}ek, C. Ahn,
K. Schoutens and A. Sevrin, {\sl `Superspace WZW models and Black
Holes'}, hep-th/9110035, Workshop on String and Related Topics,
Trieste, Itaty, Aug. 8--9 1991. Published in Trieste HEP and Cosmology,
(1991) 995.}
\lref\cft{See for example the wonderful book
 {\sl `Conformal Field Theory'}, P. di Francesco, P.~Matthieu and
 D. S\'en\'echal, Springer, 1997.}
\lref\anatomy{C. V. Johnson, {\sl`Anatomy of a Duality'}, 
Nucl. Phys. {\bf B521} (1998) 71, hep-th/9711082.}
\lref\nickmeal{{\sl `Orientifolds, Branes, and Duality of 4D Gauge Theories'},
 Nick Evans, Clifford V. Johnson and Alfred D. Shapere,
 Nucl. Phys. {\bf B505} (1997) 251, hep-th/9703210. }
\lref\ganor{O. J. Ganor and A. Hanany, 
{\sl `Small $E_8$ Instantons and Tensionless Non--critical Strings'},
Nucl. Phys. {\bf B474} (1996) 122, hep-th/9602120.}

\lref\super{C. V. Johnson, 
{\sl `Superstrings from Supergravity'}, hep-th/9804200.}
\newsec{Introduction and Summary}
\subsec{Motivations}
Recently\super, it has been shown that all of the ten dimensional
superstring theories, described in the infinite momentum frame (IMF)
by the ``matrix string'' description, have a similar qualitative
structure in the region of weak string coupling:

$\bullet$ At weak coupling they are all described by 1+1 dimensional
infra--red fixed point theories which are essentially trivial orbifold
conformal field theories. These theories may be described as the flow
from an effective 1+1 dimensional field theory: the obvious matrix
extension of the relevant Green--Schwarz action, whose prototype was
discussed in this framework in ref.\dvv.

$\bullet$ In the same limit, there is an approximate supergravity
description, dual (or nearly so) in the sense of
ref.\refs{\juan,\correspond}\ which is simply the near horizon
geometry of the fundamental string solution of a species T--dual to
the matrix string in question.

$\bullet$ The neighbourhood of the core of the supergravity solution
corresponds, {\it via} the duality map, to the weak matrix string coupling
limit. In the limit, the flow to the trivial fixed point (describing
the free matrix string) moves one to the center of the supergravity
solution, where the curvature diverges, and the dual description
breaks down, as it should.

Far away from weak coupling, the matrix descriptions of the strings
cease to all resemble  one another, and become either 0+1
dimensional (for the type~IIA or $E_8{\times}E_8$ heterotic systems)
or 2+1 dimensional (for the type~IIB or $SO(32)$ type~I/heterotic
systems). This is of course consistent with the fact that the very
strong coupling limits of all of the strings are somewhat different
from each other, according to string duality: The first two are dual to
eleven dimensional supergravity, while the latter are dual to ten
dimensional string theories.

In the case of the latter class, the natural description of the theory
at intermediate coupling is a 2+1 dimensional {\sl interacting} fixed
point theory. The theory has a supergravity dual described as eleven
dimensional supergravity compactified on $AdS_4{\times}S^7$, for the
type~IIB system, or an orbifold $AdS_4/\IZ_2{\times}S^7$ for the
$SO(32)$ system. The isometries of the compactification translate into
the superconformal symmetries and R--symmetries of the 2+1 dimensional
conformal field theory living at the boundary of $AdS_4$.

Matrix string theory is a useful alternative way of defining and
characterizing string theories. In the case of ten dimensions we now
have a complete\super\ understanding of the overall structure of these
theories, and a good understanding of when we can expect a dual
supergravity description to help in studying the defining field
theory.

There arises the obvious question: What is the analogous story for the
more newly discovered class of
superstring\refs{\stromingervafa,\dvviii,\dvv,\dvvii,\seiberg}\
theories, the ones which live in six dimensions? These theories have
certain properties which make it interesting to begin answering the
question. In the light of what was learned for the ten dimensional
theories, we can anticipate some of the structure of the
matrix--string--{\it via}--supergravity description for the six
dimensional strings:

$\bullet$ The strings all seem to be most naturally defined at
intermediate coupling. This is believed to follow from the fact that
they are self--dual objects, naturally coupled electrically and
magnetically to a three--form field strength $H^{(3)}$. For this to be
true, their coupling is frozen at some value of the coupling of order
one.

$\bullet$ This means that the strings are always interacting, and
therefore we should not expect that the matrix string theory will
involve a trivial orbifold conformal field theory. Instead, there will
be some non--trivial interacting theory. This is already known to be
true for the $(0,2)$ or ``type iia'' theory. We will see that it is
indeed true for all of the theories.

$\bullet$ We should expect further that there should exist a
supergravity dual description of the theories. This dual will be
complete in the sense that there will be no curvature singularities in
the solution, giving us a complete dual theory. For the $(0,2)$
theory, the relevant fixed point is conjectured\juan\ to be dual to
type~IIB supergravity compactified on
$AdS_3{\times}S^3{\times}T^4$. We will see that in every case, the
$AdS_3{\times}S^3$ space will arise as the dual, although (of course)
the supergravity will be different in each case.

We see therefore that the structure of the matrix string definition, or
equivalently, the supergravity origin of all of the (IMF) six dimensional
string theories is rather simple compared to the ten dimensional
theories, precisely because they prefer not to be defined at weak
coupling.

\subsec{Summary of Results}

$\bullet$ Using the defining matrix theory of longitudinal
M5--branes\berkooz, and following the appropriate limits, we observe
that the matrix strings are all defined in terms of 1+1 dimensional
interacting fixed points.  (This was already observed for the $(0,2)$
little string\refs{\dvvii,\ofereva,\edhiggs}.)

$\bullet$ The limits which define the matrix string theories also
define certain supergravity backgrounds, which can be interpreted as
``dual'' descriptions in the sense of ref.\juan.  The dual
descriptions are all smooth:

\hskip0.5cm\vbox{$\odot$ The $(0,2)$ theory is given\juan\ by type~IIB
supergravity on $AdS_3{\times}S^3{\times}T^4$.}

\hskip0.5cm\vbox{\hsize=5in $\odot$ The $(1,1)$
theory comes from type~IIA supergravity on
$AdS_3{\times}S^3{\times}T^4.$}

\hskip0.5cm\vbox{\hsize=5in $\odot$ The $(0,1)$ $E_8{\times}E_8$ theory is 
defined by $SO(32)$ heterotic supergravity on
$AdS_3{\times}S^3{\times}T^4$, or alternatively, type~IIA supergravity
on $AdS_3{\times}S^3{\times}K3$.}

\hskip0.5cm\vbox{\hsize=5in$\odot$ The $(0,1)$ $SO(32)$ theory
is defined by $E_8{\times}E_8$ heterotic supergravity on
$AdS_3{\times}S^3{\times}T^4.$}

$\bullet$ In all cases therefore, there is the appropriate $SO(2,2)$
bosonic component of the superconformal symmetry and $SO(4)$
R--symmetry. The supersymmetry of the relevant supergravity supplies
the appropriate fermionic extension. The R--symmetry has the dual
interpretation as the Lorentz group in this light--cone definition of
the little strings.

$\bullet$ In the two $(0,1)$ cases, the extra $SO(32)$ or
$E_8{\times}E_8$ global symmetries of the little heterotic string
theories\seiberg\ arise as global symmetries of their defining fixed
point theories. These in turn come from the fact that the supergravity
compactification will produce a gauge symmetry in $AdS_3$ in each
case. The AdS/CFT correspondence then demotes this gauge symmetry to a
global symmetry of the boundary theory in a similar way to what
happens for the Kaluza--Klein gauge symmetries arising from isometries
of the $S^3$.

\newsec{The case of type~iia}

We start with the matrix theory definition of M--theory in the
infinite momentum frame (IMF). It is given by\bfss\ the ${\cal N}{=}16$
supersymmetric $U(N)$ quantum mechanics arising from $N$ coincident
D0--branes' world--volume, in the limit $\ell_s{\to}0$ and
$N{\to}\infty$. The special longitudinal direction, $x^{10}$,
(initially compactified on a circle of radius $R_{10}$), is
decompactified in the limit also.  The type~IIA string theory used to
define this theory has parameters:
\eqn\params{g^{\phantom{.}}_{\rm IIA}=
R_{10}^{3/2}\ell_p^{-{3/2}},\quad\ell_s=\ell_p^{3/2}R_{10}^{-{1/2},} }
where $\ell_s$ is the string length and $\ell_p$ is the eleven
dimensional Planck length.

Our ultimate goal is to construct the six dimensional $(0,2)$
interacting string theory living on the world volume of a collection
of NS--fivebranes of the type IIA theory. Such branes originate from
M--theory as M5--branes, transverse to the circle which shrinks to
give the type~IIA string. Such branes are placed in the matrix theory
by adding hypermultiplets to the quantum mechanics, a procedure which
is really adding\berkooz\ D4--branes to the $N$~D0--brane system in
the defining type~IIA theory. Let us add $M$ such D4--branes, oriented
along the directions $x^1,\ldots,x^4$. We need to tune this
hypermultiplet theory into its Higgs branch, which is to say we
dissolve the D0--branes into the D4--branes, endowing them with $N$
units of D0--brane charge.

This system therefore defines $M$ M5--branes oriented along
$x^1,\ldots,x^4,x^{10}$, with $N$ units of momentum in the $x^{10}$
direction. Following the usual matrix string procedure, we may now
imagine that the momentum is actually along the $x^5$ direction and
shrink that direction to get a definition of the resulting type~IIA
system. In doing so, we arrive at an economical description of the
system by $T_5$--dualizing the defining type~IIA system, giving a
type~IIB string theory configuration consisting of $M$ D5--branes with
$N$ D1--branes (or a single D1--brane wound $N$ times on ${\hat x}^5$,
the dual direction).

The 1+1 dimensional Yang--Mills coupling on the D1--branes'
world--volume is given by: ${1/g_{\rm YM}^2}{=} {\ell_s^2/g_{\rm IIB}}
{=}
\ell_s^2{R_5/R_{10}}$.
As the radius $R_5$ shrinks to zero, the ten dimensional type~IIB
string coupling gets very large. We have a weakly coupled description
in terms of the S--dual system of $M$ F5--branes (a shorter term for
NS--fivebranes) with $N$ F1--branes (fundamental type~IIB strings)
inside their world volume. We shall sometimes think of this as one
F1--brane with $N$ units of winding in the ${\hat x}^5$ direction.

After $T_5$--dualizing again, we obtain a type~IIA system of $M$
F5--branes with F1--branes (fundamental type~IIA strings this time)
inside their world--volume with $N$ units of momentum in $x^5$.

This chain of dualities is similar to the chain of reasoning which
defines the matrix (IMF) ten dimensional type~IIA string\dvv: There,
the defining lagrangian came from a system of D1--branes with $N$
units of winding. This was S--dual to a system of wound type~IIB
F1--branes. A T--duality on the winding direction gave the type~IIA
string (F1--brane) with $N$ units of momentum. The Fock space of the
IMF matrix string was built up from these winding type~IIB strings,
and the explicit description at all couplings was given in terms of
the D1--brane system, which is a 1+1 dimensional Yang--Mills theory: a
matrix--valued type~IIA Green--Schwarz action.  At weak coupling, the
target space of the theory (moduli space of the 1+1 dimensional
Yang--Mills theory) is simply\refs{\banks,\dvv}\
$S^N(\IR^8){\equiv}(\IR^8)^N/S_N$, where $S_N$ is the group of
permutations of $N$ objects, the D1--branes, and $\IR^8$ is the space
allowed D1--brane positions, the permitted values of the (in general
matrix--valued) bosonic fields of the Yang--Mills theory. This is an
orbifold theory. As shown in ref.\refs{\motl,\banks,\dvv}, the twisted
sectors of this orbifold describe long type~IIA strings which can
survive in the large $N$ limit to define strings with finite momentum
in the IMF direction\foot{Ref.\dvv\ went on to show how to switch on
interactions in this second quantized definition of the type~IIA
string.}.

The same thing happens here. There is a non--trivial interacting
theory living on the type~IIA F5--brane's world--volume even as we
take the limit $g_{\rm IIA}{\to}0$, as argued in ref.\seiberg. The
``little strings'' (sometimes called\refs{\dvvii,\dvviv}\
``microstrings'') which carry the basic degrees of freedom of the
theory are described by the 1+1 dimensional theory we have defined. It
is a 1+1 dimensional theory derived from the D1--branes' +
wrapped\foot{We will generically think of the five--branes as wrapped
on a $T^4$, transverse to the strings. Therefore they also contribute
to the stringy 1+1 dimensional model. $N$ and $M$ are now on similar
footing, and our earlier identification of $N$ as the momentum of the
little strings is modified by $M$, as wrapping induces some more
D1--brane charge proportional to $M$. The schematics of the discussion
is correct for motivational purposes, as the reasoning in this paper
will not need the full discussion of the
momentum\refs{\dvvii,\dvviii,\dvviv}.}\ D5--branes' world--volume. The
theory has $M$ hypermultiplets in the fundamental of $U(N)$ and it has
been tuned to its Higgs branch. In other words, as instantons of the
D5--brane's $SU(M)$ gauge theory, the D1--branes are far from the
point--like limit\edsmall\ and are instead fat instantons, having
finite scale size. They are delocalized inside the D5--branes, in the
directions $\{x^1,\ldots,x^4\}$.

Furthermore, we are interested in the zero coupling limit of the final
type~IIA theory and so we should take the strong coupling limit of
this configuration. The 1+1 dimensional Yang--Mills theory is
therefore strongly coupled in the limit that we want, and
it flows to the infra--red. The resulting infra--red fixed point
defines for us the matrix $(0,2)$ ``little string theory''.

The target space of this theory is\refs{\stromingervafa,\dvvii}\ a
hyperK\"ahler deformation of $S^{NM}(T^4)$. There has been much work
devoted to this theory in the
literature\refs{\stromingervafa,\juaniv,\martinecii,\dvvii,\ofereva,
\edhiggs,\oferseiberg,\sethi}.

\subsec{The role of Type IIB Supergravity}

Notice that like the ten dimensional case, we are led to describe the
long strings in the theory as winding type~IIB F--strings.  These long
strings arise in the large $N$ limit, which we must take to properly
define the original matrix M--theory, and here in order to obtain the
light cone type~IIA string theory.

In this large $N$ limit, we must take seriously the supergravity
fields generated by the D1--brane configuration. If we take large $M$
also, we can fully describe the supergravity fields with a metric
valid for low curvature everywhere\andygaryjuan:
\eqn\donedfive{\eqalign{ds^2=&
\biggl(1+{g_{\rm IIB}\ell_s^2N\over vr^2}\biggr)^{-1/2}
\biggl(1+{g_{\rm IIB}\ell_s^2M\over r^2}\biggr)^{-1/2}\left[(-dt^2+dx_5^2)
+\biggl(1+{g_{\rm IIB}\ell_s^2M\over
r^2}\biggr)\sum_{i=1}^4dx_i^2\right]\cr +&\biggl(1+{g_{\rm
IIB}\ell_s^2N\over vr^2}\biggr)^{1/2}
\biggl(1+{g_{\rm IIB}\ell_s^2M\over r^2}\biggr)^{1/2}(dr^2+r^2d\Omega^2_3).}}

(Here, $v$ is a dimensionless measure of the volume of the $T^4$ on
which the D5--brane is wrapped.) In the limit, this solution becomes
simply $AdS_3{\times}S^3{\times}T^4$, where the radius of the first
two factors is set by the product $MN$, and the latter by $M/N$.
\eqn\adsthree{ds^2\sim\ell_s^2 \sqrt{NM}\left(u^2(-dt^2+dx_5^2)+
{du^2\over u^2}+d\Omega_3^2\right)+\sqrt{M\over N}\sum_{i=1}^4
dx_i^2.}  We have absorbed some inessential constants into $r$ and
then set $u=r/\ell_s^2$.  This theory is conjectured\juan\ to be the
dual of the 1+1 dimensional $(4,4)$ superconformal field theory we are
interested in the strong coupling IR limit. (The string coupling has
also diverged in the limit, and we need to consider the S--dual
configuration for weak coupling. This is same metric for that S--dual
IIB supergravity solution, however, as the dilaton is constant in this
limit. This structure will persist in the later discussions too.) The
theory lives at the ``boundary'' of the $AdS_3$. The required
$SO(2,2)$ superconformal algebra with its $SO(4)$ R--symmetry arises
from the isometries of the $AdS_3$ and the $S^3$ respectively. This
theory has been studied from this point of view recently in
refs.\refs{\boon,\deger,\juanandy,\martinec,\jan,\vafaii}.

This AdS/CFT correspondence is conjectured to be the full description
of the non--trivial conformal field theory. In this sense, the $(0,2)$
``little string'' theory (in the infinite momentum frame) has a
supergravity origin.

\newsec{The case of type~iib}

There is a little string theory living on F5--branes of type~IIB
string theory as well. We should try to characterize it also.

Starting again with our matrix definition of M5--branes, we may
proceed to descend to type~IIB theory, with its F5--branes by
compactifying on an additional circle, $x^4$, in addition to the one
which we shrunk to get the type~IIA theory. We are shrinking M--theory
on a torus, and therefore should obtain the type~IIB
theory\refs{\paul,\john}. The extra detail of the $M$ D4--branes in
our defining type~IIA theory should be interesting.

In doing so, we obtain, after $T_{45}$--dualizing, a type~IIA string
theory again with $M$ D4--branes located now in $\{x^1,x^2,x^3,{\hat
x}^5\}$ with $N$ D2--branes in $\{{\hat x}^4,{\hat x}^5\}$. These
D2--branes are delocalized inside the D4--branes, as before.  They are
not infinite in the ${\hat x}^4$ directions, as the D4--branes are
pointlike there, and so they end on them. We get the directions
$\{{\hat x}^4,{\hat x}^5\}$ directions both decompactified when we
define the resulting type~IIB theory at intermediate coupling. To get
a weakly coupled type~IIB string theory, we would let ${\hat x}^5$
grow faster than ${\hat x}^4$. In a frame where we fix ${\hat x}^5$
large, we see that the ${\hat x}^4$ direction shrinks away. We shall
do this presently.

The effective gauge coupling of the 1+1 dimensional theory living on
the part of the D2--brane is given by:
${1/g_{\rm YM}^2}{=}{\ell_s/{\tilde g}_{\rm IIA}}
{=}{R_4R_5/R_{10}}$.
For small $R_4,R_5$, both the dual type~IIA string theory coupling
${\tilde g}_{\rm IIA}$ and the Yang--Mills coupling $g_{\rm YM}$ are
large. This means that we should consider our configuration as an
M--theory configuration: eleven dimensional supergravity with
branes. The $M$ D4--branes become $M$ M5--branes, now stretched along
$\{x^1,x^2,x^3,{\hat x}^5,x^{10}\}$, while the $N$ M2--branes are
stretched along $\{{\hat x}^5,{\hat x}^4\}$. They end on the
M5--branes, and are delocalized inside them.

The weakly coupled type~IIB string limit, with $M$ F5--branes, is
described by taking ${\hat x}^4{\to}0$, giving $M$ F5--branes in
type~IIA with fundamental ${\hat x}^5$--wound F--strings inside;
$T_5$--duality completes the route to the type~IIB system with
F--strings possessing momentum in $x^5$.  Requiring weakly coupled
type~IIB therefore focuses the discussion on M5--brane world--volume.
The leg of the M2--branes not inside the M5--branes becomes less
important to the discussion and the physics of the effective string
inside the M5--branes' worldvolume dominates. This 1+1 dimensional
theory therefore describes whatever interacting theory there is on the
F5--brane at weak type~IIB string coupling.

\subsec{The role of Eleven Dimensional Supergravity}

Furthermore, the large $N, M$ limit allows us to discuss the system in
terms of the supergravity solution\refs{\tseytlin,\gauntlett}:
\eqn\mtwomfive{\eqalign{ds^2=&
\biggl(1+{\ell_p^2N\over vr^2}\biggr)^{-1/3}
\biggl(1+{\ell_p^2M\over r^2}\biggr)^{-2/3}(-dt^2+dx_5^2)\cr
+&\biggl(1+{\ell_p^2N\over vr^2}\biggr)^{-1/3}
\biggl(1+{\ell_p^2M\over r^2}\biggr)^{1/3}(dx_1^2+dx_2^2+dx_3^2
+d{\hat x}_{10}^2)\cr
+&\biggl(1+{\ell_p^2N\over vr^2}\biggr)^{1/3}
\biggl(1+{\ell_p^2M\over r^2}\biggr)^{-1/3}d{\hat x}_4^2\cr
+&\biggl(1+{\ell_p^2N\over vr^2}\biggr)^{2/3}
\biggl(1+{\ell_p^2M\over r^2}\biggr)^{1/3}(dr^2+r^2d\Omega^2_3).}}

(Here, $v$ is a dimensionless measure of the volume of the $T^4$ on
which the M5--brane is wrapped. $r^2{=}\sum_{i{=}6}^9x_i^2$) In the
limit, this solution becomes\mads\ simply
$AdS_3{\times}S^3{\times}T^4{\times}S^1$. It is easy to compute that
(after a rescaling) the radius of the first two factors is set by the
product $(MN^2)^{1/3}$:
\eqn\madsthree{\eqalign{ds^2\sim&(MN^2)^{1/3}\ell_p^2
\left(u^2(-dt^2+d{\hat x}_5^2)+
{du^2\over u^2}+d\Omega_3^2\right)\cr+&\left({M\over
N}\right)^{1/3}(dx_1^2+dx_2^2+dx_3^2+d{\hat x}_{10}^2)+\left({N\over
M}\right)^{1/3}d{\hat x}_4^2.}}  (Again, we have absorbed some
constants into $r$ and then set $u{=}r/\ell_p^2$.) This is the dual
supergravity description of the theory in the limit. Notice that we
get this eleven dimensional supergravity solution at $R_4{=}R_5$,
which means that the type~IIB coupling we are studying is not small,
but at the self--dual value $g_{\rm IIB}{=}R_5/R_4{=}1$. We actually
want the limit $g_{\rm IIB}{\to}0$, if we are to directly study the
decoupling limit which yields the physics of the little string trapped
inside the F5--brane.

\subsec{The Role of Type~IIA Supergravity}
We therefore need to study this geometry in the ten dimensional limit
that ${\hat x}^4{\to}0$. Luckily, as the ${\hat 4}{\hat 4}$ metric
component is $(N/M)^{1/3}$, a constant, there is no resulting
non--trivial dilaton dependence for the ten dimensional theory, and no
need to multiply the rest of the metric in eqn.\madsthree\ by any
functions of the radial variable.

Using the relation between the ten and eleven dimensional
metrics\refs{\town} ($A$ is the R--R one--form potential):
\eqn\relations{ds_{11}^2=
e^{4\phi/3}\left[(d{\hat x}_4+A^\mu
dx_\mu)^2+e^{-2\phi}ds_{10}^2\right],} it is easily established that
our metric \madsthree\ becomes precisely the ten dimensional solution
\adsthree.  We are therefore left with type~IIA supergravity\foot{Yes,
we could have arrived at this from a more direct starting point:
T--dualizing the type~IIB $AdS_3$ compactification in the $T^4$,
therefore constructing a type~IIA solution. It was nevertheless
instructive to proceed by this route, seeing where the matrix
prescription takes us, in the spirit of ref.\super.}\ compactified on
$AdS_3{\times}S^3{\times}T^4$. It is natural to conjecture that the
AdS/CFT correspondence defines for us a 1+1 dimensional superconformal
field theory on the boundary with the correct superconformal algebra
and R--symmetries as before. Of course, the details of the theory are
different, as they should be: This is a different supergravity theory.
 
This 1+1 dimensional superconformal field theory defines the $(1,1)$
six dimensional little string theory. This fixed point has a dual
supergravity solution which is smooth everywhere. The type~iib system
in the infinite momentum frame therefore arises from a simple
supergravity description.

\newsec{The case of the little $E_8{\times}E_8$ heterotic string.}
The next step is obvious. We may place a family of $M$ M5--branes into
the $E_8{\times}E_8$ string theory by introducing $M$ D4--branes into
the defining D0--brane system, which additionally contains 16
D8--branes and 2 O8--planes\foot{For a reminder of the details of this
starting point, see ref.\super. The original references are
refs.\refs{\danielsson,\eva,\banksmotl\matrixheterotic,\rey}.}. We
orient the eight dimensional objects in
$\{x^1,\ldots,x^4,x^6,\ldots,x^9\}$, and the D4--branes in
$\{x^1,\ldots,x^4\}$, as before. This defines M--theory on an interval
(in $x^5$) defined by an M9--plane at each end of it, with $M$
M5--branes located pointwise along it. Everything has momentum in the
$x^{10}$ direction.

As usual, we can choose to place the momentum in the $x^5$ direction,
and shrink it. The theory becomes a type~IIB system with $M$
D5--branes, with $N$ D1--branes delocalized inside them. The
background of 16 D9--branes has an $SO(16){\times}SO(16)$ Wilson line.

The (0,4) 1+1 dimensional theory on the world volume of the D1--branes
has $M$ hypermultiplets from the 1--5 sector and 32 fermions from the
1--9 sector. Without the D5--branes, this theory goes in the strong
coupling limit to an IR fixed point which defines the weakly coupled
$E_8{\times}E_8$ heterotic string. In the present case, the flow
defines the content of the interacting (0,1) six dimensional theory on
the world--volume of the F5--brane of the $E_8{\times}E_8$ heterotic
string theory. This interacting theory has also a global
$E_8{\times}E_8$ symmetry. (See also refs.\matrixeight, for related
models.)

\subsec{The role of $SO(32)$ Supergravity}
In similar fashion to that which we described for the type~iia system,
the large $N$ and $M$ limit tells us to examine the supergravity
fields around the D1--D5 system, but now in type~IB string theory.

The supergravity solution is precisely the same as it was for the
type~IIB case. The D5--branes remain small instantons of the D9--brane
gauge group and so there is no modification to the supergravity
solution by an expression for large instantons.

Therefore, in the limit we are led to type~IB's $SO(32)$ supergravity
compactified upon $AdS_3{\times}S^3{\times}T^4$, and in the strong
coupling limit (implied by shrinking $R_5$) we replace this with the
heterotic $SO(32)$ supergravity with the same compactification (this
is valid as the dilaton is constant). Formally, we still need to have
winding around the ${\hat x}^5$ direction. {\it Near} the limit, we
can think of it as a large circle, and the spacetime become $AdS_3$ in
the limit. We can place the required $SO(16){\times}SO(16)$ Wilson
line around this direction, as dictated by the model.

The near--$AdS_3$ inherits a gauge symmetry $SO(16){\times}SO(16)$
from ten dimensions. There are states in the adjoint
$\bf(120,1){+}(1,120)$. We expect that in the presence of the Wilson
line, the $N{\to}\infty$ approach to $AdS_3$ will give masses to
states in the $\bf(16,16)$, while states in the $\bf(128,1){+}(1,128)$
become massless, performing the expected enhancement to
$E_8{\times}E_8$ at strong coupling\foot{We have no direct proof of
this, but we will supply an indirect one using string duality in the
next section. It will be interesting to find a direct demonstration of
this phenomenon.}

It is natural to conjecture that there is a non--trivial 1+1
dimensional superconformal field theory living at the boundary of the
$AdS_3$ space. This is the matrix theory of the $E_8{\times}E_8$
little heterotic string theory, with infinite momentum frame in the
$x^5$ direction. In this $N{\to}\infty$ limit, the correct long
strings should emerge in the usual way.  The gauge symmetry
anticipated above gives rise to a global $E_8{\times}E_8$ symmetry of
the interacting conformal field theory living at the boundary, and
hence of the six dimensional spacetime little string theory.

\subsec{The role of Type~IIA/Heterotic Duality}
There is of course another route to defining the $E_8{\times}E_8$
little heterotic string. This may also be viewed as a proof of the
correctness of the above prescription (particularly the incomplete
argument for the gauge/global symmetry) using type~IIA/heterotic
duality\refs{\goed,\harveystrominger,\sendual}.

In section~4.1 we recovered the $SO(32)$ type~IB system compactified
on $AdS_3{\times}S^3{\times}T^4$. As a ten dimensional system, we took
it to strong coupling. We used ten dimensional $SO(32)$
type~IB/heterotic duality to relate this to a weakly coupled heterotic
system compactified on the same space.

Next, we can consider the system as six dimensional again and replace
the $T^4$ with a $K3$ while replacing the heterotic theory with the
type~IIA theory. As shown in ref.\harveystrominger, the type~IIA
string recovers the $E_8{\times}E_8$ symmetry of a dual heterotic
string from the intersection lattice of the $K3$.

So simply replacing the $T^4$ by a $K3$ in the type~IIA supergravity
$AdS_3{\times}S^3{\times}T^4$ compactification already established in
section~3.2, we get the $E_8{\times}E_8$ little heterotic string
theory.

This way of realizing the string sharpens our earlier discussion of
the origin\foot{This is also in line with the intuition (expressed to
me by T. Banks and H. Verlinde) that once we have compactified to six
dimensions, the heterotic string should be best understood in terms of
type~IIA on $K3$. See also ref.\paulii\ for how this reasoning may be
pushed back up all the way to eleven dimensions.}\ of the required
$E_8{\times}E_8$ global symmetry, while the duality to the type~IB
system of the previous subsection points to its correctness.

\newsec{The case of the little $SO(32)$ heterotic string}
To define the $SO(32)$ system, we start with the D0--D4--D8--O8 theory
from the previous section and shrink $x^4$ as well as $x^5$.  As in
the case of type~IIB, we arrive at the case of intermediate coupling
for the parent theory, if we treat both directions the same way:
$R_4{\sim}R_5$.

Taking the same limits as section~3, we arrive at an M--theory
configuration involving $N$~M2--branes stretched along ${\hat x}^4$,
ending on $M$~M5--branes which are pointlike along the ${\hat x}^4$
interval defined by the two M9--planes.

To get the limit of weak heterotic $SO(32)$ coupling, we would send
the size of the ${\hat x}^4$ interval to zero, and the resulting
strongly coupled 1+1 dimensional theory of the endpoints of the
M2--branes inside the M5--branes is the theory we want.  In the weakly
coupled heterotic limit, where the size of the interval shrinks away
completely, these 1+1 dimensional endpoints become heterotic
F1--strings inside heterotic F5--branes. 

In taking the limit, we have not tuned things such that the M5--branes
stay away from the M9--plane endpoints even in the limit. This would
give expectation values to some number of tensor multiplets in the
resulting six dimensional theory. Nor have we tuned things such that
the F5--branes dissolve into the M9--planes, becoming fat
instantons. Instead, we are just at the dividing edge between these
branches of the F5--brane moduli space, so that the full
$E_8{\times}E_8$ gauge symmetry is present; the F5--branes are small
instantons, and we are able to disappear down the infinite throat to
find the decoupled theory. This is the interacting six dimensional
theory we seek.

Let us take the large $N,M$ limit, as before. We can then define what
the supergravity dual should be.

\subsec{The role of Eleven Dimensional Supergravity}
At intermediate coupling for the $SO(32)$ system, it is easy to see
that the large $N,M$ limit of the M2--M5--M9 system becomes eleven
dimensional supergravity compactified upon
$AdS_3{\times}S^3{\times}T^4{\times}S^1/\IZ_2$, the boundaries of the
orbifolded ${\hat x}^4$ direction are the M9--planes. They are
infinitely far apart in the limit $R_4{=}0$. 

This should be contrasted with the case of the pure $SO(32)$ system at
intermediate coupling encountered in ref.\super. There, the system was
defined by $AdS_4/\IZ_2{\times}S^7$, {\it i.e.,} the orbifold acted in
the $AdS_4$ giving fixed points inside the space\foot{Interestingly, a
quick calculation shows that the geometry of the 9+1 dimensional fixed
point in that case is (after rescaling using \relations\ to measure in
ten--dimensional units) precisely the near--horizon geometry of a
fundamental string.}.

Here, the system is much simpler, as the orbifold misses the $AdS$
component entirely, due to the presence of the M5--branes.

As before, we actually want the weak $SO(32)$ heterotic coupling
($g_{\rm HB}{=}R_5/R_4$) limit, which corresponds to shrinking the
${\hat x}^4$ direction. There is no radial dependence of the metric
component in this direction, and so the resulting ten dimensional
theory is also simple (and non--singular). The relevant supergravity
theory is of course now the $E_8{\times}E_8$ heterotic
supergravity\horavawitten.

\subsec{The role of $E_8{\times}E_8$ Supergravity}
In the limit, we have $E_8{\times}E_8$ supergravity compactified on
$AdS_3{\times}S^3{\times}T^4$. Formally, near the limit we still need
to have winding around the ${\hat x}^5$ direction, and so We can think
of it as an extremely large circle which we will asymptote to the
$AdS_3$ in the limit. We then place an $SO(16){\times}SO(16)$ Wilson
line in this direction.

In the $N{\to}\infty$ limit, (and $R_5{\to}0$ limit) the long strings
emerge in the usual way. We expect also that the correct
$SO(16){\times}SO(16)$ quantum numbers survive to this time recover
$SO(32)$ symmetry instead of $E_8{\times}E_8$, consistent with the
known $T_5$--duality with the Wilson line. Again, we have no direct
proof of this mechanism.

It is natural to conjecture that there is a non--trivial 1+1
dimensional superconformal field theory living at the boundary of the
$AdS_3$ space. This is the matrix theory of the $SO(32)$ little
heterotic string theory, with infinite momentum in the $x^5$
direction.

\newsec{Closing Remarks and Outlook}
So we see that, in the spirit of ref.\super, interacting matrix string
theories are captured by smooth dual supergravity compactifications
involving $AdS$. 

The four little string theories are
naturally interacting six dimensional theories obtained by
``capturing'' ten dimensional strings with F5--branes\seiberg. 

This is of course why they are four in number: only four of the ten
dimensional superstring theories can ensnare such descendents, as the
pure type~IB system does not contain the requisite F5--branes with
which to do the capturing. (In some sense, it also does not contain an
honest defining type~IB string either: There is no NS--NS $B$--field.)

This is all consistent with what we observe here, because no $AdS_3$
geometry arises in the limit of weak type~IB coupling. This is because
the legs of the M2--brane defining the intermediate coupling situation
(see section~5.1) are not on the same footing (pun intended) in this
case\foot{One should contrast with the type~IIB matrix situation
involving $AdS_4$ in ref.\super. There, both legs of the M2--brane are
inside the $AdS_4$.}. One sees that the weak type~IB coupling limit
(obtained by shrinking ${\hat x}^5$) would lead back to type~IA
supergravity (as in ref.\super), but the $AdS_3$ gets spoiled. At
best, this leads to a 0+1 dimensional theory with a singular
supergravity limit: presumably a non--interacting theory defined by a
quantum mechanics with a trivial orbifold moduli space, following the
philosophy of the present paper and ref.\super.

The four theories have all been shown to have supergravity defining
duals which involve $AdS_3{\times}S^3$. In each case, the supergravity
changes appropriately to give the correct fermionic extension to
$SO(2,2)$ to fill out the required supersymmetry algebra for the
defining 1+1 dimensional fixed point theory on the boundary. 

In the heterotic cases the supergravity also supplies the required
gauge symmetry, although we did not supply a direct argument for how
$SO(32)$ and $E_8{\times}E_8$ get exchanged from an $AdS$ supergravity
point of view: Somehow, a more careful examination of the interplay
between the Wilson line and the approach to the $AdS_3{\times}S^3$
geometry (where the circle goes away) should give the required result
that the $SO(32)$ system (plus Wilson line) gives $E_8{\times}E_8$
gauge symmetry in the limit and {\it vice--versa}. It is an $AdS$
supergravity analogue of T--duality with Wilson lines. The
type~IIA/heterotic duality argument presented in section~4.2 is so far
the best direct supergravity argument presented here.

At risk of over--emphasizing the point, let us remark upon the
simplicity of the overall structure we have uncovered here for all of
the string theories, combining the results of ref.\super\ and the
present paper:

$\bullet$ For the five (IMF) ten dimensional superstring theories near
weak coupling, there is a dual description in terms of a solution of
the supergravity associated to the T--dual species of string. The
solution is merely the near horizon geometry of the fundamental string
solution in ten dimensions.

$\bullet$ The theories have weak coupling limits where they become
free. This is represented by the fact that the fundamental string
solution is singular at the core: supergravity must break down there
as it cannot describe such a trivial theory.

$\bullet$ For the four (IMF) six dimensional superstring theories,
there is a dual supergravity solution, again in terms of the
supergravity of the T--dual species of string. This time, the solution
is simply the near horizon geometry of the fundamental string inside
the six dimensional world--volume of a NS--fivebrane.

$\bullet$ These F1--F5 solutions are perfectly smooth everywhere. This
is consistent with the fact that the dual little strings seem to have
no weak coupling limits.
 
A final remark to be made is about the current discrepancy observed
between the spectrum of supergravity on $AdS_3$ and that of 1+1
dimensional conformal field theories\vafaii. The structure of our
observations is not affected by this. We have not compared spectra here,
but only supersymmetry and global symmetries. We expect that the
AdS/CFT correspondence for 1+1 dimensional superconformal field
theories will be at the {\it very} least a useful guide, where at
least some of the structures in supergravity and conformal field
theory organize one another. The extent to which the supergravity
captures all of the physics, including correlators\martinec, {\it
etc.}, remains to be seen.

The structures uncovered here may be regarded as added motivation for
trying to gain understanding of the AdS/2DCFT correspondence, as it
will give  a handle on the little string theories, which are
certainly going to be important in the final story.

\bigskip

\noindent
{\bf Acknowledgments:}

\noindent
CVJ was supported in part by family, friends and music. CVJ gratefully
acknowledges useful conversations with Tom Banks, Jan de Boer, Mike
Douglas, Emil Martinec, Robert C. Myers, Amanda Peet and Herman Verlinde. This
research was supported financially by NSF grant
\#PHY97--22022.

\bigskip
\bigskip

\centerline{\epsfxsize1.0in\epsfbox{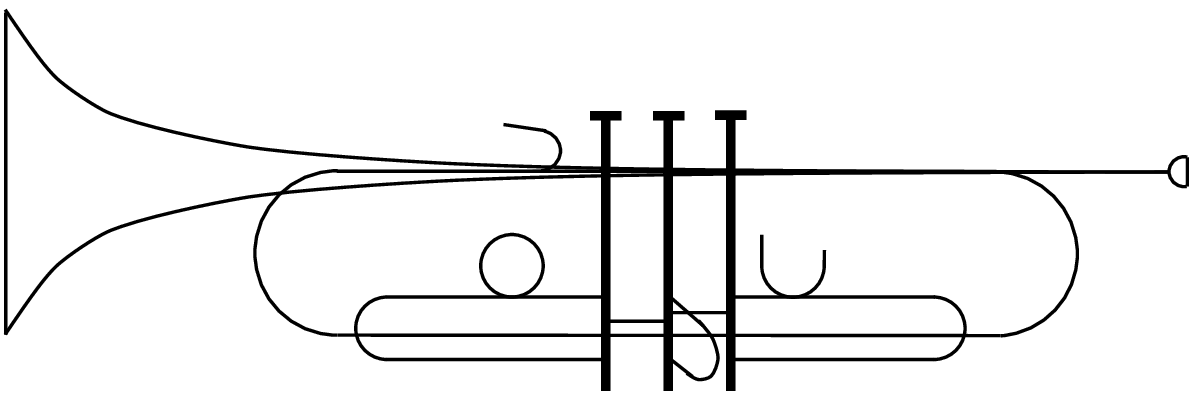}}

\listrefs

\bye